\documentclass[final,10pt,twocolumn,letterpaper,conference]{IEEEtran}

\usepackage{amsmath,amsfonts,amssymb,amsbsy,url,verbatim}
\usepackage{mathtools}
\usepackage{times}
\usepackage[english]{babel}
\usepackage[dvips]{graphicx}
\usepackage{bm}
\usepackage{comment}
\usepackage{stfloats}
\usepackage{cite}
\usepackage{psfrag}
\usepackage{dsfont} 
\usepackage{url}
\usepackage[top=0.75in, bottom=1.05in, left=0.65in, right=0.65in]{geometry}

\newcommand{\bi}{\begin{itemize}}
\newcommand{\ei}{\end{itemize}}
\newcommand{\ben}{\begin{enumerate}}
\newcommand{\een}{\end{enumerate}}
\newcommand{\bc}{\begin{cases}}
\newcommand{\ec}{\end{cases}}
\newcommand{\bd}{\begin{description}}
\newcommand{\ed}{\end{description}}

\newcommand{\be}{\begin{equation}}
\newcommand{\ee}{\end{equation}}
\newcommand{\bea}{\begin{eqnarray}}
\newcommand{\eea}{\end{eqnarray}}

\newcommand{\back}{\!\!\!\!\!}

\graphicspath{%
     {./figures/}%
}

\IEEEoverridecommandlockouts

\begin{document}

\title{Towards an Appropriate Beamforming Scheme for Initial Cell Discovery in mmW 5G Cellular Networks \vspace{-4mm}}

%An Energy Consumption Based Comparative Analysis of Analog, Digital and Hybrid Beamforming Schemes for
%Initial Cell Discovery in mmW Cellular Networks

\author{%
    \IEEEauthorblockN{Waqas~bin~Abbas, Michele Zorzi}%
    \IEEEauthorblockA{Department of Information Engineering, University of Padova --- via Gradenigo 6/B, 35131 Padova, Italy \vspace{0.5mm}}%
    \IEEEauthorblockA{E-mail: \texttt{\small \{waqas,zorzi\}@dei.unipd.it}}%
}

\maketitle

\begin{abstract}
    Beamforming is an essential requirement to combat
high pathloss and to improve signal-to-noise ratio during initial
cell discovery in future millimeter wave cellular networks. The
choice of an appropriate beamforming is directly coupled with
its energy consumption. 
The energy consumption is even of more concern at a battery limited mobile station (MS). In this work, we provide an energy consumption based comparison of different beamforming schemes
while considering both a low power and a high power analog-to-digital converter (ADC) for a millimeter wave based receiver at the MS. We
analyze both context information (CI) (GPS positioning based)
and non context information based schemes, and show that
analog beamforming with CI (where mobile station's positioning information is already available) can result in a lower energy
consumption, while in all other scenarios digital
beamforming has a lower energy consumption than analog and hybrid beamforming. 
We also show that under certain scenarios recently proposed phase shifters network architecture can result in a lower energy consumption than other beamforming schemes.
Moreover, we show that the energy consumption trend among different beamforming schemes is valid irrespective of the number of ADC bits.
Finally, we propose a new signaling structure which utilizes a relatively higher frequency sub-carrier for primary synchronization signals compared to other signaling, which allows a further reduction in initial cell search delay and energy consumption of the MS.  
\end{abstract}

% \begin{IEEEkeywords}
% \noindent Underwater networks, selective repeat ARQ, multiuser transmissions, simulation, WOSS, parameter optimization.
% \end{IEEEkeywords}

\section{Introduction}
\label{sec:intro}

At millimeter wave (mmW) huge bandwidths are available
to fulfill the data rate requirement of the next generation
(5G) cellular networks \cite{KhanF_mmWave}. 
However, unlike in microwave bands, mmW
communication suffers from high path-loss \cite{5G_WillWork}. To address
this issue, spatial beamforming is considered as an inevitable
choice for mmW communication \cite{Roh_mmWaveBF}.
Unlike in LTE, beamforming (BF) in mmW communication is not only an essential requirement for data plane communication but also required during control plane signaling.
BF is even a more critical requirement during the initial cell discovery phase, when the received signal power is often well below the noise floor.

In addition to BF, Low latency is also one of the fundamental requirements for 5G \cite{Andrews_5Gbe}. Low latency can be achieved by incorporating a high bandwidth sub-carrier.
However, this increase in the bandwidth requires higher sampling frequency for an analog-to-digital converter (ADC) which corresponds to a higher power consumption. This power consumption and latency product is even more critical during initial cell discovery (due to large search delays) and directly dictates the choice of an appropriate BF scheme (especially at the MS's receiver design).

Analog, Digital and Hybrid beamforming (ABF, DBF and HBF) are  
the commonly discussed BF schemes for mmW communication \cite{SM_BF_2014}.
%In mmW communication, the utilization of huge bandwidth B along with a higher number of antennas results in a high power consumption, therefore.
The choice of an appropriate BF scheme is critically related to its power consumption,
%Typically, power consumption of analog-to-digital converter ADC increases linearly with B, which makes the choice of a low power BF scheme,
and it is even of more importance for a mmW receiver design especially at the mobile station (MS) which has limited available battery. 
In general, DBF requires a separate RF chain for each antenna element and therefore has the highest power consumption, whereas ABF consumes the least power. 
However, DBF allows to form multiple simultaneous beams to cover the whole angular space at a particular instant, whereas ABF only converges to a single beam at a particular time instant. 
Therefore, ABF results in a larger time delay than DBF or HBF due to an additional time required to form beams in different directions to cover the whole angular space.
This shows a trade-off among the power consumption and the directional search delay, which is critical during initial cell discovery (ICD). 
In this work, we study this trade-off and identify the appropriate BF scheme at the MS for ICD.

\subsection{Related Work}
\label{ssec:RW}
The research related to directional cell discovery in mmW 5G cellular networks is very recent.
The authors in \cite{RC_mmW_14} suggested to scan the complete angular space sequentially to identify the best BF direction both at the MS and the base station (BS).
In \cite{Barati_IBF}, directional cell discovery is studied, and the authors showed that DBF with a low-bit ADC at the MS can be preferable compared to the ABF.
A delay based comparison for initial access in mmW cellular networks is studied in \cite{Barati_IA_16}, where it is shown that DBF has a lower delay than ABF without any performance degradation.
In \cite{Alkhateeb_IBF}, the advantages of HBF in terms of lower delay and better access probability than ABF are presented for the case of initial beamforming.

To address the issue of large search delay regarding the identification of the right BF direction, a context information (CI) based directional cell search is proposed in \cite{CaponeFS15_CI_BF}.
The authors consider a HetNet scenario where the CI about the MS positioning is forwarded to the mmW BS and then the BS transmits the initial synchronization signals in the provided direction.
In \cite{myCI_PSN16}, to reduce the directional search delay associated with the ABF, the authors consider the availability of the CI regarding the mmW BS positioning at the MS.
They further proposed a phase shifters network (PSN) architecture (which results in a lower power consumption than HBF) to mitigate the effect of erroneous CI.

In \cite{OrhanER15_PowerCons}, the total power consumption based comparison of an ABF and a DBF for the mmW receiver at the MS is studied. 
Results showed that with low power ADC and a fixed total power consumption budget, DBF may result in a higher capacity than ABF when the channel state information is available at the transmitter.
The authors in \cite{My_PCComp_EW16} compare the total power consumption of the ABF, DBF and HBF based receiver architectures considering both a low power and a high power ADC model, and discuss the feasibility of different beamforming schemes for both the MS and the BS and for both data plane and control plane signaling.
However, to the best of our knowledge there is no previous work which discusses the choice of an appropriate BF scheme (at the MS's receiver) for ICD considering both the power consumption and the time delay (latency) together for both the non-CI and the CI based scenarios.

In this work, focusing on initial cell discovery (ICD) for the CI and the non-CI based scenario, we study this power consumption vs time delay trade-off and identify the choice of an appropriate BF scheme based on the total energy consumption.

%\cite{KhanF_mmWave},\cite{5G_WillWork}
\subsection{Our Work}
\label{ssec:OW}

In this work, we compare the receiver energy consumption associated with different BF schemes during initial
cell discovery. Primarily, we focus on an energy consumption comparison among the
ABF, DBF, HBF
and PSN  receiver architectures at the MS. Energy consumption is computed as the product of power
and time delay. Power consumption for different beamforming schemes is computed considering similar mmW receiver
architectures as in \cite{My_PCComp_EW16}, while delay is calculated following a synchronization signaling structure as used in LTE. For power
consumption, we consider both a low power ADC (LPADC)
and a high power ADC (HPADC) model. We also consider both
context information (CI) and non CI based scenarios for
initial cell discovery. In CI based initial cell discovery, we
consider that the GPS coordinates of the mmW base
stations (BS) position are provided to the MS
by the microwave BS. We divide the CI based scenario
into 1) where the GPS coordinates of the MS’s positioning
are already available, and 2) where the MS has to acquire
its GPS coordinates by using assisted GPS (AGPS) which
introduces an additional delay. % and therefore energy consumption compared to the former CI based scenario.
Moreover, we only consider CI based scenario for AFB, HBF
and PSN, as DBF already allows to form multiple beams simultaneously to cover the complete angular space, and therefore CI for DBF will not result in any notable advantage. In the rest of the paper, we refer to non CI based scenario
as nCI, CI with GPS coordinates already available, i.e., CI
with no delay as CInD, and CI with the delay of acquiring the
GPS coordinates as CID. Considering the above mentioned
scenarios, we show that

\begin{itemize}
\item All BF schemes have a similar time delay for CInD scenario, whereas in other scenarios DBF has a minimum delay.
\item Energy consumption (EC) of all beamforming schemes decreases with an increase in the sub-carrier bandwidth.
\item ABF (for both LPADC and HPADC) only results in
a lower EC with a CI based scheme
where the GPS coordinates regarding MS’s positioning
are already available.
\item In all scenarios except CInD, DBF results in a lower
EC than ABF and HBF.
\item PSN has a lower EC than other BF schemes for higher sub-carrier bandwidth.
\item The EC trends among different BF schemes are independent of the number of ADC bits.
\end{itemize}

Finally, we propose a new signaling structure for 5G, where the primary synchronization signals are generated with a higher sub-carrier bandwidth while all other signaling is performed at a lower bandwidth. 
This scheme allows to cover the angular space by the BS with a relatively short delay than with lower sub-carrier bandwidth, and therefore also reduces the energy consumption.  

\section{Initial Cell Discovery}
\label{sec:ICD}

Initial cell discovery (ICD) procedures proposed for conventional cellular networks are relatively different from what
is required for mmW cellular communications. Conventionally, the transmission and the reception of the primary synchronization
signals (PSS) and secondary synchronization signals (SSS) (essential for ICD and synchronization) at the BS
and the MS, respectively, is done omnidirectionally. 
However, in
mmW cellular networks, directional communication is essential even during the ICD process to exploit the BF gain. 
In addition to beamforming, reduced latency is another important requirement for 5G cellular
networks and this can be achieved by increasing the bandwidth
of the sub-carrier $B_{SC}$. 
Note that this increase in $B_{SC}$ will also increase the total system bandwidth $B_{Tot}$. 
Although an increase in $B_{Tot}$ reduces the latency, it also results in a higher consumption at the mmW receiver, mainly because the ADC needs to sample the incoming signal at a higher rate.
%Firstly, this increase in $B_{SC}$ directly
%corresponds to an increase in ADC power consumption, and it
%significantly increases the difference among the power consumption
%of a fully digital beamforming receiver (i.e., with DBF) and
%an analog beamforming receiver (i.e., ABF). Secondly, an
%increase in $B_{SC}$ also reduces the symbol time, which corresponds
%to a reduction in the total reception delay. 
%Moreover, DBF results in a least ICD delay as it fewer cycles to scan the whole angular space as compared to other beamforming schemes. 
We study this power consumption and time delay relationship
for different BF schemes.
We next discuss the LTE control signaling model to evaluate the
synchronization signaling time delay.

\begin{figure}
    \centering
    \psfrag{Energy Consumption}{Energy Consumption}
    \psfrag{BW in MHz}{$B_{SC}$}
    \psfrag{nCI HPADC}{nCI HPADC}
    \includegraphics[width=\columnwidth]{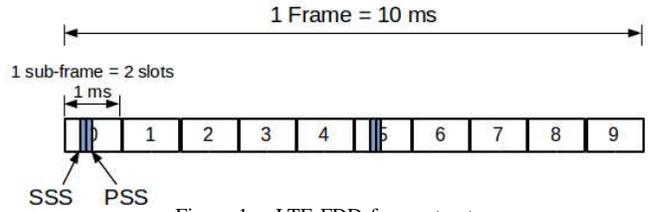}\vspace{-6.5mm}
    \caption{LTE FDD frame structure.}
    \label{fig:PSS}
\end{figure}

\subsection{Initial Cell Discovery in LTE}
\label{ssec:ICD_LTE}
To initialize the ICD procedure, the MS starts looking for
the reception of the synchronization signals (i.e, PSS and SSS).
These synchronization signals provide the initial synchronization and also identify the physical layer cell identity.
There are 504 different physical layer cell identities, where the
SSS has 168 sequences and provides the cell identity group
and the PSS has 3 different
sequences and provides identity within the cell identity group. In
FDD, the PSS and the SSS are transmitted once every
5 ms for a duration of 71.4 $\mu$s (Figure \ref{fig:PSS}). Both synchronization signals
constitute 6 resource blocks (corresponds to 72 sub-carriers)
with 62 sub-carriers centered around DC (irrespective of the operating bandwidth of LTE) while the rest 10 sub-carriers are unused. After the successful reception of the PSS
and the SSS, the MS finishes the initial cell discovery and
starts the initial access process.

\subsection{Extension to mmW}
\label{ssec:ICD_mmW}
Unlike in LTE, in mmW 5G cellular networks an additional requirement for initial cell discovery is to identify the required
beamforming direction to improve the SNR at the receiver. In comparison to LTE, this identification of the beamforming direction results in an additional ICD delay. This delay
depends on the size of the antenna array and on the beamforming
scheme incorporated both at the MS and at the BS. In this work,
we consider 64 antennas at the BS and 16 antennas at the MS,
which with beamspace beamforming requires $64 \times 16 = 1024$
different directional scans ($N_{D}$) to cover the complete angular space.
The total ICD delay is a combination of the delay associated
with 1) the control signaling (PSS and SSS) and 2) the
identification of the beamforming direction.

  \begin{table}
     \centering
     \caption{Synchronization Signal Time Period and Total System Bandwidth ($B_{Tot}$) Associated to Different $B_{SC}$}\vspace{-1mm}
     \begin{tabular}{|c||c|c|c|c|c|}
         \hline
         $B_{SC}$ (MHz) &  0.015 (LTE) & 0.25 & 0.5 & 1 & 10\\ 
         \hline
        $T_{PSS}$  & 5 ms & 0.3 ms & 0.15 ms & 75 $\mu$s & 7.5 $\mu$s \\
         \hline
        $B_{Tot}$ (MHz)  & 1.4 & 23.3 & 46.7 & 93.3 & 933.3 \\
         \hline
     \end{tabular}
     \label{tab:TPss}
 \end{table}

\section{Energy Consumption in mmW ICD}
\label{sec:EC_mmWICD}

We measure the energy consumption for different BF schemes considering the three different scenarios
discussed in Section \ref{ssec:OW}. 
As Energy consumption (EC) is computed as
the product of power consumption and time delay, we first compute the time delay associated with nCI, CInD and CID, whereas the power consumption of the different BF schemes will be studied in the following subsections.

%   &  \begin{tabular}{@{}c@{}}$t_{Del}$ (nCI) \\ ($S_C \times T_{PSS}$) \end{tabular} & \begin{tabular}{@{}c@{}}$t_{Del}$ (CInD) \\ ($S_C \times T_{PSS}$) \end{tabular} & \begin{tabular}{@{}c@{}}$t_{Del}$ (CID) \\ ($S_C \times T_{PSS}$) \end{tabular}

\subsection{Time Delay}
\label{ssec:TD}
We follow the LTE signaling structure to calculate the delay
associated with the reception of synchronization signals, and
identify the delay corresponding to different values of the sub-carrier
bandwidth $B_{SC}$. This delay corresponds to the time period associated
to the transmission of the PSS signal, $T_{PSS}$, for different ICD
scenarios, as shown in Table \ref{tab:TPss}. To receive the initial synchronization signals, the MS has to look in each BF direction for
a duration of $T_{PSS}$. Therefore, the total time delay $t_{Del}$ can
be computed as the product of $T_{PSS}$ and $N_{D}$. The total time
delay corresponding to different BF schemes and different ICD
scenarios is shown in Table \ref{tab:TD}. We consider only ABF at the
BS while the MS can form its combining beam using either
ABF, DBF, HBF or PSN. This shows that the BS at a particular
instant can transmit the synchronization signal only in one
angular direction, therefore the BS requires 64 (equal to the number
of antennas) different transmissions at separate time instants to cover the
whole angular space. Moreover, the angular search delay at the MS is related to
the available beamforming scheme. 

  \begin{table}
     \centering
     \caption{$t_{Del}$ for Different BF Schemes Corresponding to Different ICD Scenarios, with ABF at the BS}\vspace{-1mm}
     \begin{tabular}{|c||c|c|c|}
		\hline
          &  $t_{Del}$ (nCI) & $t_{Del}$ (CInD) & $t_{Del}$ (CID)\\          
         \hline
         ABF  &  $1024 \times T_{PSS}$ & $64 \times T_{PSS}$ & $64 \times T_{PSS} + t_{CI}$\\ 
         \hline
        DBF  & $64 \times T_{PSS}$ & $64\times T_{PSS}$ & $64 \times T_{PSS}$ \\
         \hline
        HBF   & $256 \times T_{PSS}$ & $64 \times T_{PSS}$ & $64 \times T_{PSS} + t_{CI}$\\
         \hline
          PSN   & $256 \times T_{PSS}$ & $64 \times T_{PSS}$ & $64 \times T_{PSS} + t_{CI}$\\
         \hline
     \end{tabular}
     \label{tab:TD}
 \end{table}

With ABF, the MS has to look in 16 different angular directions, which results in $N_{D} = 64\times 16 = 1024$, the maximum scan search delay. Unlike ABF, $N_{D}$ for HBF (PSN) is directly related
to the number of available RF chains (combiners). In this
work, the number of RF chains and combiners for HBF and
PSN is set to 4. This allows both architectures to form
4 simultaneous beams in different angular directions, and
therefore $N_{D} = 64 \times 16/4 = 256$ for both HBF and PSN. Finally, DBF only requires
64 directional scans to cover the complete angular space. The above
mentioned $N_{D}$ corresponds to the nCI scenario, and $t_{Del}$ can
be computed as the product of $N_{D}$ and $T_{PSS}$ (Table \ref{tab:TD}). This $t_{Del}$ nCI showed that DBF has the least delay whereas ABF results in maximum delay.

We consider the availability of CI only for ABF, HBF and
PSN, as the fully digital architecture in DBF allows to form beams
in all angular directions simultaneously, and therefore CI will
not provide any significant advantage with DBF receivers. The
availability of CI is advantageous for ABF, HBF and PSN as it reduces the number of directional scans. With the availability of CI with no
delay (i.e., CInD), the receiver already has the knowledge of
the appropriate combining direction and therefore the number of directional scans for ABF, HBF and PSN with CInD is similar to that for DBF
(i.e., 64). This shows that $t_{Del}$ for CInD for all beamforming
schemes is similar.

Finally, a scenario where the GPS coordinates of the MS
are not available (i.e., CID) results in an additional time delay.
To acquire the GPS coordinates, we considered an AGPS
protocol in which the time delay $t_{CI}$ corresponding to the
acquisition GPS coordinates is generally communication channel dependent. However, to simplify the analysis we consider the Time to first fix (TTFF) with AGPS to be approximately around 1 s (as mentioned in \cite{van_AGPS}), and  assume $t_{CI} = 1.5$ s, slightly more than TTFF. This
is a relatively short delay, however, we show that even this
short $t_{CI}$ results not only in a higher delay but also in a relatively
higher energy consumption compared to the other
design choices.

  \begin{table}
     \centering
     \caption{Receiver Power Consumption for Different BF Schemes for HPADC}\vspace{-1mm}
     \begin{tabular}{|c||c|c|c|c|}
		\hline
          $B_{SC}$ &  DBF & ABF & HBF & PSN\\          
         \hline
         15 KHz  & 1.31  & 1  & 2.43  & 2.49\\ 
         \hline
        250 KHz  & 1.87  & 1.03  & 2.57  & 2.52 \\
         \hline
        500 KHz & 2.57  & 1.07  &  2.72 & 2.56\\
         \hline
          1 MHz & 3.67  & 1.15  & 3.02  & 2.63\\
         \hline
          10 MHz  & 25.1616  & 2.4891  & 8.369  & 3.978\\
         \hline
     \end{tabular}
     \label{tab:HPADC}
 \end{table}

   \begin{table}
     \centering
     \caption{Receiver Power Consumption for Different BF Schemes for LPADC}\vspace{-1mm}
     \begin{tabular}{|c||c|c|c|c|}
		\hline
          $B_{SC}$ &  DBF & ABF & HBF & PSN\\          
         \hline
         15 KHz  & 1.28  & 0.996  & 2.43  & 2.3\\ 
         \hline
        250 KHz  & 1.3  & 0.998  & 2.43  & 2.3\\
         \hline
        500 KHz  & 1.32  & 0.999  & 2.44  & 2.3\\
         \hline
          1 MHz  & 1.37  & 1  &  2.45 & 2.31\\
         \hline
          10 MHz & 2.22  & 1.06  & 2.66  & 2.36\\
         \hline
     \end{tabular}
     \label{tab:LPADC}
 \end{table} 
 
\subsection{Power Consumption}
\label{ssec:PC}

Receiver power consumption is an important concern due
to the utilization of huge bandwidths at mmW communication.
In this work, we consider the total power consumption of
the receiver. The power consumption of all other components
except ADC is considered independent of the total system bandwidth. We consider two different power consumption models
for an ADC, i.e., a low power and a high power ADC, to better
quantify the energy consumption of different beamforming
schemes at these two power extremes. For details regarding
the receiver architecture and power consumption for
different beamforming schemes the reader is referred to \cite{myCI_PSN16}, \cite{My_PCComp_EW16}.

The total system bandwidth (as shown in Table \ref{tab:TPss}) is an
important parameter, as an increase in $B_{Tot}$ linearly increases the ADC power consumption (as $B_{Tot}$ defines the required sampling rate of an ADC).
$B_{Tot}$ is directly proportional to the sub-carrier bandwidth. An increase in $B_{SC}$ results in a higher $B_{Tot}$ and
therefore a higher ADC power consumption.
$B_{Tot}$ for different $B_{SC}$ is computed following the LTE
format used for the 1.4 MHz bandwidth. 
With 1.4 MHz of bandwidth, 6 RBs are transmitted at each time instant.
A RB consists of 12 sub-carriers with $B_{SC}$ = 15 KHz.
This corresponds to a bandwidth of 1.08 MHz for 6 RBs, and therefore with total bandwidth of 1.4 MHz, the utilized bandwidth is 77.14\% (i.e., $1.08$ MHz $\times$ $100/1.4$ MHz $= 77.14\%$). 
$B_{Tot}$ for different $B_{SC}$ can be computed similarly and is listed in
Table \ref{tab:TPss}.

The power consumption (in Watts) of different beamforming
corresponding to different $B_{SC}$ for a HPADC and a
LPADC is listed in Table \ref{tab:HPADC} and Table \ref{tab:LPADC}, respectively. 
The power consumption values are obtained for a 6 bits ADC.
The listed values show that the rate of increase in the power
consumption is related not only to an increase in $B_{Tot}$ but also to
the choice of the ADC design. For instance, the rate at which
power consumption increases for HPADC is much higher as
compared to LPADC. Moreover, DBF with LPADC results in
a lower power consumption than HBF and PSN even for
$B_{Tot} = 10$ MHz.

Finally, the energy consumption (EC) can be computed by
evaluating the product of the total power consumption and the
time delay. % which we will discuss in the next section.

\section{Results}
\label{sec:Res}
\begin{figure}
    \centering
    \psfrag{Energy Consumption}{\back\back\back \ Energy Consumption (Joules)}
    \psfrag{BW in MHz}{\ \ \ $B_{SC}$}
    \psfrag{nCI HPADC}{\back\back\back\back nCI HPADC (10 bits ADC)}
    \scalebox{0.95}[0.70]{\includegraphics[width=\columnwidth]{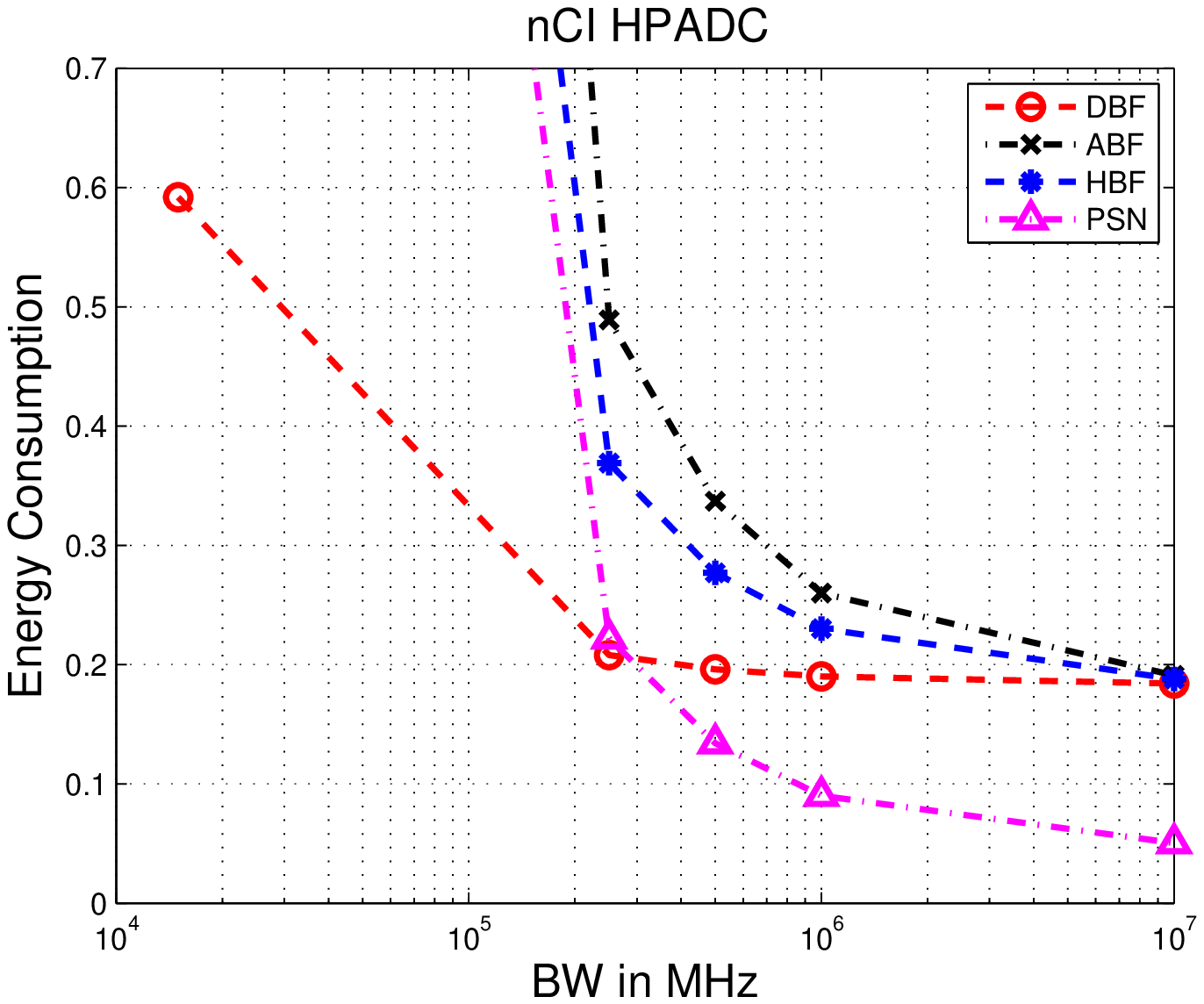}}\vspace{-3mm}
    \caption{Energy consumption vs $B_{SC}$ for different beamforming schemes with nCI scenario and HPADC.}
    \label{fig:nCI_HPADC10b}
\end{figure}

We now discuss the energy consumption trend followed
by different BF receiver architectures during the ICD, and
identify the choice of the appropriate BF scheme for nCI, CInD
and CID scenarios while considering both a low power and
a high power ADC model. The energy consumption of different
BF schemes considering different $B_{SC}$ is computed using the
power consumption and time delay values listed in Tables \ref{tab:HPADC}, \ref{tab:LPADC}
and \ref{tab:TD}, respectively.
All results except those in Figure \ref{fig:nCI_HPADC10b} (where we considered 10 bits) are obtained for a 6 bits ADC.
%Moreover, Figures \ref{fig:nCI_HPADC10b} and \ref{fig:nCI_HPADC} show that the trend among different BF schemes is valid irrespective of the number of ADC bits.

\begin{figure}
    \centering
    \psfrag{Energy Consumption}{\back\back\back \ Energy Consumption (Joules)}
    \psfrag{BW in MHz}{\ \ \ $B_{SC}$}
    \psfrag{nCI HPADC}{nCI HPADC}
    \scalebox{0.95}[0.70]{\includegraphics[width=\columnwidth]{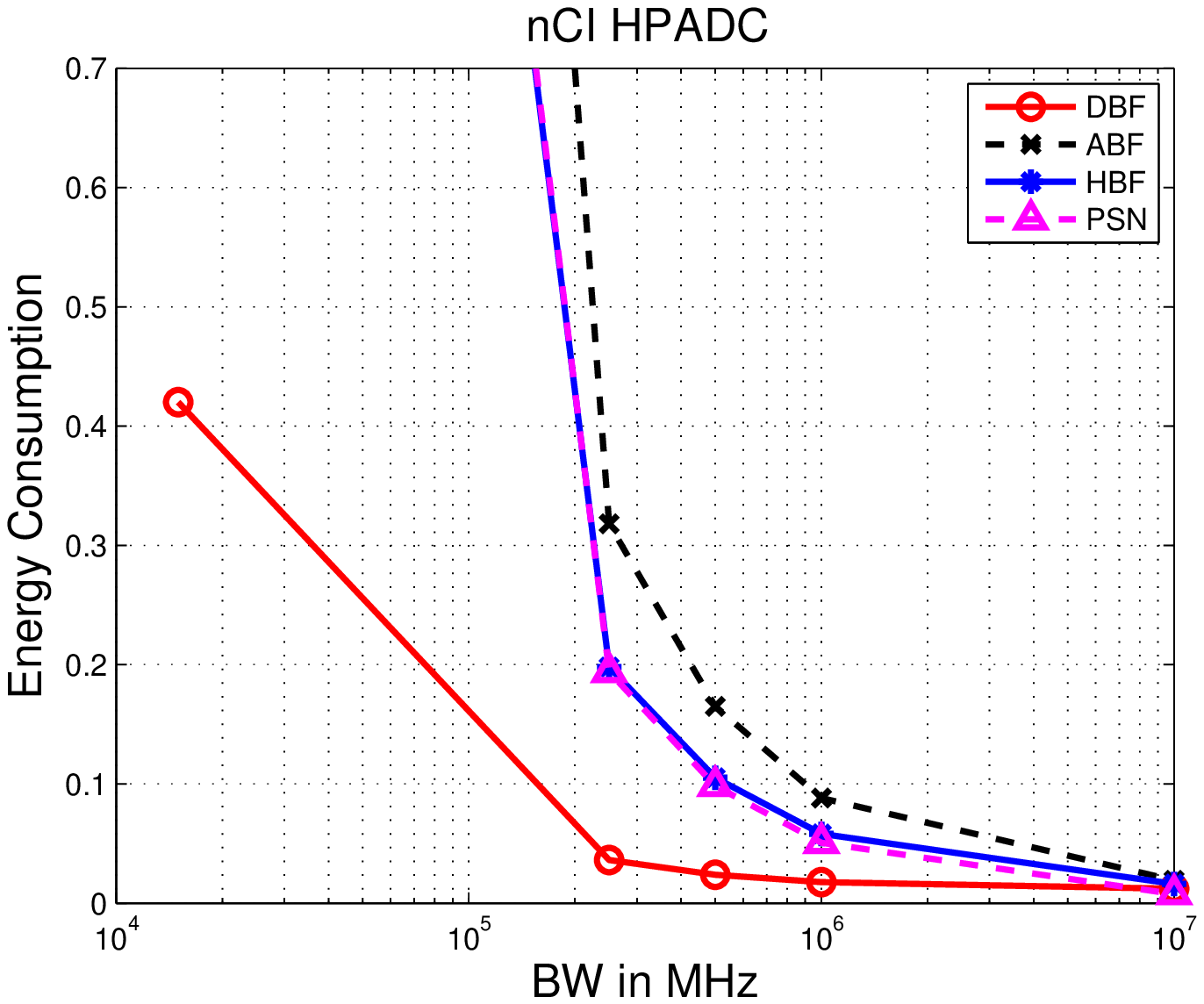}}\vspace{-3mm}
    \caption{Energy consumption vs $B_{SC}$ for different beamforming schemes with nCI scenario and HPADC.}
    \label{fig:nCI_HPADC}
\end{figure}

\begin{figure}
    \centering
    \psfrag{Energy Consumption}{\back\back\back \ Energy Consumption (Joules)}
    \psfrag{BW in MHz}{\ \ \ $B_{SC}$}
    \psfrag{nCI HPADC}{nCI LPADC}
    \scalebox{1}[0.70]{\includegraphics[width=\columnwidth]{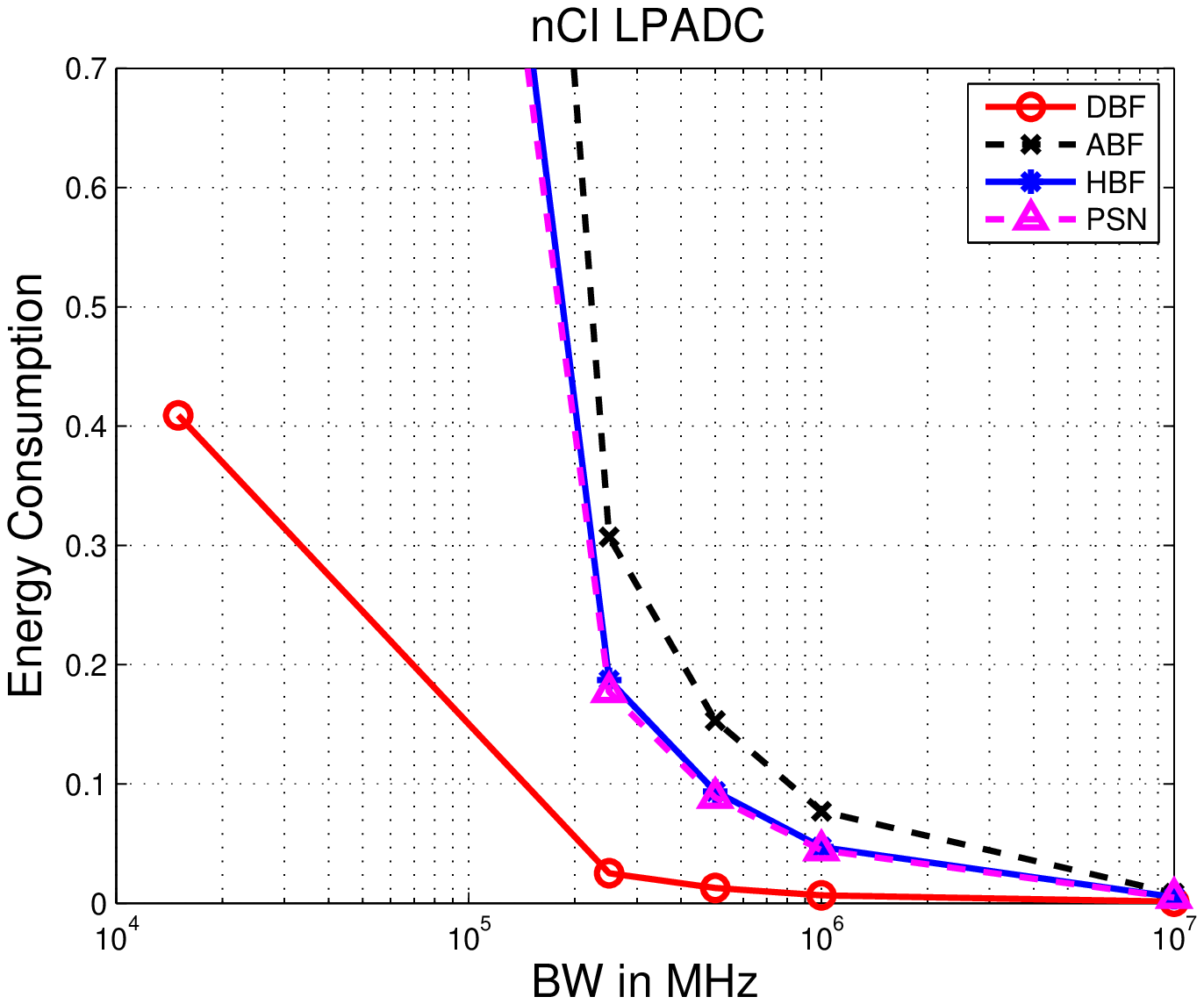}}\vspace{-3mm}
    \caption{Energy consumption vs $B_{SC}$ for different beamforming schemes with nCI scenario and LPADC.}
    \label{fig:nCI_LPADC}
\end{figure}

\begin{figure}
    \centering
    \psfrag{Energy Consumption}{\back\back\back \ Energy Consumption (Joules)}
    \psfrag{BW in MHz}{\ \ \ $B_{SC}$}
    \psfrag{CInD HPADC}{CInD HPADC}
    \scalebox{1}[0.70]{\includegraphics[width=\columnwidth]{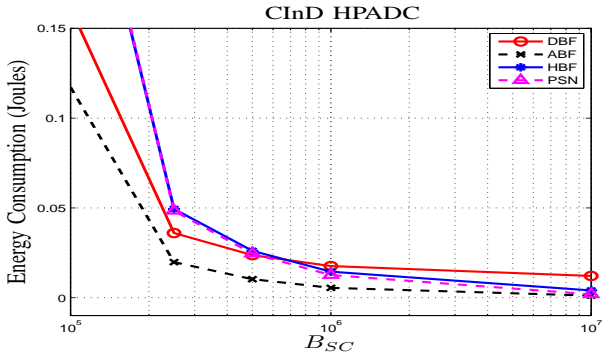}}\vspace{-3mm}
    \caption{Energy consumption vs $B_{SC}$ for different beamforming schemes with CInD scenario and HPADC.}
    \label{fig:CInD_HPADC}
\end{figure}

\begin{figure}
    \centering
    \psfrag{Energy Consumption}{\back\back\back \ Energy Consumption (Joules)}
    \psfrag{BW in MHz}{\ \ \ $B_{SC}$}
    \psfrag{CInD HPADC}{CInD LPADC}
    \scalebox{1}[0.70]{\includegraphics[width=\columnwidth]{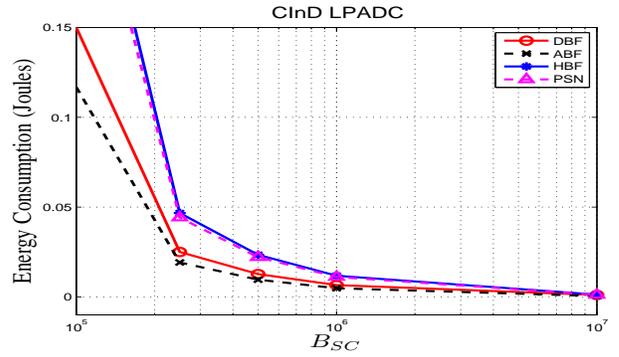}}\vspace{-3mm}
    \caption{Energy consumption vs $B_{SC}$ for different beam with CInD scenario and LPADC.}
    \label{fig:CInD_LPADC}
\end{figure}

\begin{figure}
    \centering
    \psfrag{Energy Consumption}{\back\back\back \ Energy Consumption (Joules)}
    \psfrag{BW in MHz}{\ \ \ $B_{SC}$}
    \psfrag{CID HPADC}{CID HPADC}
    \scalebox{1}[0.70]{\includegraphics[width=\columnwidth]{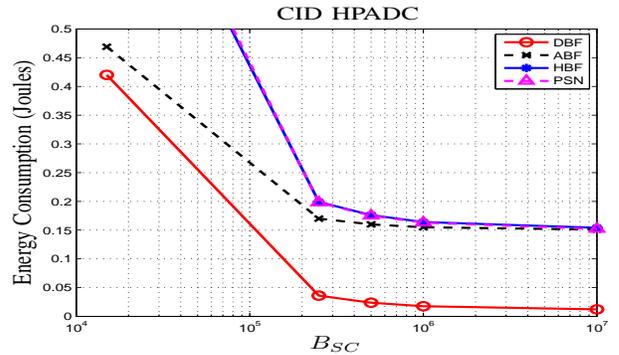}}\vspace{-3mm}
    \caption{Energy consumption vs $B_{SC}$ for different beam with CID scenario and HPADC.}
    \label{fig:CID_HPADC}
\end{figure}

\begin{figure}
    \centering
    \psfrag{Energy Consumption}{\back\back\back \ Energy Consumption (Joules)}
    \psfrag{BW in MHz}{\ \ \ $B_{SC}$}
    \psfrag{CID LPADC}{CID LPADC}
    \scalebox{1}[0.70]{\includegraphics[width=\columnwidth]{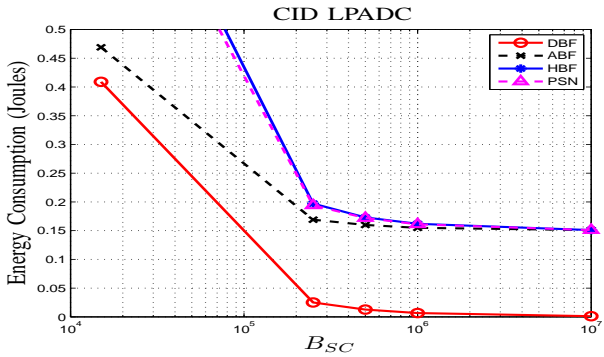}}\vspace{-3mm}
    \caption{Energy consumption vs $B_{SC}$ for different beam with CID scenario and LPADC.}
    \label{fig:CID_LPADC}
\end{figure}

Figures \ref{fig:nCI_HPADC10b}, \ref{fig:nCI_HPADC} and \ref{fig:nCI_LPADC} show the energy consumption vs $B_{SC}$
for the nCI scenario. The plots show that the energy consumption
of all BF schemes decreases (with a decreasing slope) with
an increase in $B_{SC}$\footnote{ the plots of Figures \ref{fig:CInD_HPADC}-\ref{fig:CID_LPADC} also show a similar trend i.e., decrease in the EC
with an increase in the $B_{SC}$}. This follows from the fact that $t_{Del}$
decreases with an increase in $B_{SC}$, therefore the EC of the receiver components other than ADC decreases. However, the EC of an ADC remains
constant as the power consumption of the ADC increases with the
same rate with which $t_{Del}$ decreases. The plots also show
that, even with a single RF chain, ABF results in a higher
EC relative to the other beamforming for both
the HPADC and the LPADC model. This shows that $t_{Del}$
for ABF is large enough (due to higher $N_{D}$) to dominate 
the lower power consumption of the ABF which results in 
the highest EC for ABF. On the other hand, interestingly, 
DBF always has a lower EC than ABF and HBF, irrespective of the number of ADC bits and of the ADC power model. 
However, the difference among the EC of DBF, ABF and HBF decreases with an increase in $B_{SC}$ and starts converging to a similar EC. 
We will discuss this convergence later. 

The EC trend followed by PSN is relatively different than what is observed for other BF schemes.
Firstly, PSN has a higher energy consumption than DBF at lower $B_{SC}$, but results in a lower EC than DBF at higher $B_{SC}$.
The converging value of the EC of PSN is also low compared to other BF schemes.
This is because the PSN architecture requires ADCs similar to ABF but with much fewer $N_{D}$.
Secondly, the intersection point between DBF and PSN shifts towards lower $B_{SC}$ with an increase in the number of ADC bits.
This is because an increase in bits results in a higher ADC power, due to which the ADC power starts dominating the power consumption of the other components, which results in a rapid convergence in the EC of PSN.
Finally, from Figures \ref{fig:nCI_HPADC10b} and \ref{fig:nCI_HPADC} it is obvious that the trend among different BF schemes is valid irrespective of the number of ADC bits. 
 
% for a LPADC, whereas 
%with a HPADC model DBF still has a minimum EC except 
%for $B_{SC} = 10$ MHz where PSN has a slightly lower EC than 
%DBF. Moreover, PSN has a lower EC than HBF at higher $B_{SC}$ 
%and vice versa. These results show that for the scenario where 
%CI is not available, DBF with its lower energy may be a valid 
%receiver architecture even with a HPADC model. 

Figures \ref{fig:CInD_HPADC} and \ref{fig:CInD_LPADC} show an EC comparison among different 
BF schemes for a CInD scenario. In this scenario, it is 
considered that the GPS coordinates of the MS are already 
available and therefore all BF schemes have a similar $t_{Del}$. The plots show 
that ABF has the lowest EC in this scenario. This is due to 
the reduction in $N_{D}$ thanks to the availability of CI regarding 
the direction of the mmW BS. Also note that, despite the
availability of CI, DBF has a lower or similar EC compared to
PSN and HBF up to $B_{SC} = 10$ MHz and $B_{SC} = 1$ MHz for a
LPADC model and a HPADC, respectively. The trend among
HBF and PSN is similar to what was observed for the nCI scenario.
Although these results depict the feasibility of ABF for ICD
considering the CInD scenario, to better analyze the feasibility of ABF it is also important to note
that the performance of ABF degrades with an increase in the
error in the available CI, where PSN has a more robust performance (as shown in \cite{myCI_PSN16}). 
Therefore the choice of the appropriate BF scheme for the CInD scenario not only depends on the minimum EC but is also important to quantify the performance of a particular BF scheme in the case of erroneous CI.

Figures \ref{fig:CID_HPADC} and \ref{fig:CID_LPADC} 
show the EC vs $B_{SC}$ for a CID scenario.
To compute the additional EC corresponding to the 
acquisition of the GPS coordinates of the MS, the time delay is
considered to be 1.5 s (as mentioned in Section \ref{ssec:TD}) and the
power consumption is assumed to be 100 mW. The results show
that ABF has a lower EC than HBF and PSN for
small values of $B_{SC}$, but the EC of ABF, HBF and PSN is higher compared to what observed for nCI and CInD scenarios.
This is because the EC associated to the acquisition of GPS coordinates is
high enough that all BF schemes result in a higher EC.
Moreover, DBF (for both HPADC and LPADC model) has a lower EC than other schemes for all values of $B_{SC}$. 
Therefore, the CID scenario not only results
in a higher EC for ABF, HBF and PSN but also introduces an
extra time delay. %, which makes it not a feasible scenario for ICD.

\begin{figure}
    \centering
    \psfrag{EC Convergence value}{\back\back\back \ Energy Consumption (Joules)}
    \psfrag{BW in MHz}{$B_{SC}$}
    \psfrag{CID LPADC}{CID LPADC}
    \scalebox{1}[0.70]{\includegraphics[width=\columnwidth]{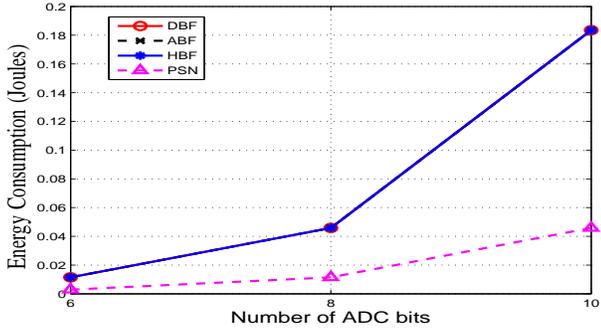}}\vspace{-4mm}
    \caption{Energy consumption convergence for different BF schemes.}
    \label{fig:EC_ConVal}
\end{figure}

\subsection{EC Convergence}
\label{ssec:ECCon}
A common trend that can be observed in Figures \ref{fig:nCI_HPADC10b}-\ref{fig:CID_LPADC} is the convergence of the EC with an increase in $B_{SC}$.
This is because the time delay is higher at lower $B_{SC}$, which makes the EC of all components other than the ADC more significant.
However, with an increase in $B_{SC}$ the EC of the ADC starts dominating.
Therefore, the convergence value for the nCI scenario (convergence for CID and CInD can be computed similarly) can simply be calculated by evaluating the EC of the total number of ADCs present in a particular BF scheme.
The EC of a single ADC can be computed as  
\begin{equation}
   E_{ADC} = P_{ADC}\times t_{Del} = cBR \times t_{Del}
    \label{eq:ECadc}
\end{equation}
where $E_{ADC}$ and $P_{ADC}$ are the energy and power consumption of an ADC, $c$ is the energy consumption per conversion and $R$ is the resolution of an ADC.
Equation (\ref{eq:ECadc}) shows that the EC of an ADC remains constant irrespective of $B_{SC}$, as an increase in $B_{SC}$ will decrease $t_{Del}$ and vice-versa. % and therefore the EC of an ADC remains unchanged.
The convergence value of a particular BF scheme can be found by computing the product $E_{ADC}$ and the number of ADC %where the $t_{Del}$ is used to compute $E_{ADC}$ and the number of ADCs is selected
based on a particular BF scheme.
The results for EC convergence are shown in Figure \ref{fig:EC_ConVal}, where the convergence value is shown for different numbers of ADC bits while considering ABF, DBF, HBF and PSN schemes.
The results show that PSN has the minimum value of the EC convergence for any number of ADC bits, whereas all other beamforming schemes converge to the same value.
This is because the product of $N_{D}$ and the number of ADC for ABF, DBF and HBF has a similar result.
However, a PSN architecture which allows to form multiple simultaneous beams and compare them in the analog domain requires a single RF chain (i.e., similar number of ADCs as used for ABF) and can cover the complete angular space with fewer directional scans than ABF, and therefore PSN results in a lower EC.
%thus PSN requires fewer scan cycles to cover the complete angular space with number of ADC similar to ABF which results in its lower EC.    

\section{Higher Sub-Carrier B for PSS Signaling}
\label{sec:HSC}
%The results in Section \ref{sec:Res} showed that with an increase in $B_{SC}$ the EC decreases.
%Based on this decrease in EC with an increase in $B_{SC}$, we propose a new signaling structure for 5G where the PSS are generated using a relatively higher sub-carrier bandwidth compared to other signaling.
Usually, control plane signaling requires a lower bandwidth compared to data plane signaling. 
For instance, in LTE the reserved bandwidth for synchronization signaling is limited to 1.4 MHz, whereas the total system bandwidth can be as large as 20 MHz based on the data rate requirement of the mobile user. 
The lower bandwidth definitely reduces the power consumption. 
However, as already shown, an increase in the sub-carrier frequency not only reduces the initial cell search delay but also the EC.  
By combining the above two points i.e., 1) the difference in the bandwidth of data plane and control plane signaling and 2) a higher sub-carrier bandwidth results in a lower latency and EC, we now propose a new signaling structure for 5G cellular networks, where PSS utilizes a relatively higher sub-carrier bandwidth compared to other signaling. 
The maximum increase in the PSS sub-carrier bandwidth ($B_{SC,PSS}$) can
be set to the maximum available bandwidth for data plane.
We next discuss the proposed signaling structure and briefly discuss its performance, whereas a detailed analysis is left as future work.

\begin{figure}
    \centering
    \psfrag{Energy Consumption}{Energy Consumption}
    \psfrag{BW in MHz}{$B_{SC}$}
    \psfrag{nCI HPADC}{nCI HPADC}
    \scalebox{1}[0.75]{\includegraphics[width=\columnwidth]{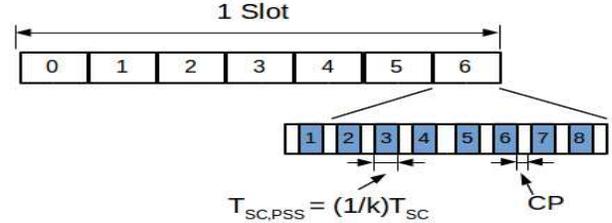}} \vspace{-7mm}
    \caption{Proposed slot structure for PSS.}
    \label{fig:HBsc}
\end{figure}

The proposed slot structure for 5G containing PSS is shown in Figure \ref{fig:HBsc}, where $T_{SC,PSS}$ and $T_{SC}$ represents the time period of the higher bandwidth PSS sub-carrier and sub-carrier bandwidth (i.e., $B_{SC}$) used for other signaling, respectively, CP represents the cyclic prefix interval corresponding to $T_{SC,PSS}$ and $k$ is a constant which represents the fractional increase or decrease in $B_{SC}$ or $T_{SC}$ for the PSS.
In Figure \ref{fig:HBsc}, $k$ is set to 8, and therefore in the time duration $T_{SC}$, 8 primary synchronization signals can be transmitted.
This is because, according to the proposed structure, $B_{SC,PSS}$ is increased $k$ times, which reduces the transmission time required for PSS by a factor of $k$ compared to what is required for sub-carriers with bandwidth $B_{SC}$.
The allows to transmit multiple $B_{SC,PSS}$ signals in either a different or the same angular direction in a similar time interval as that of other signaling, i.e., $T_{SC}$.
This has two main advantages, 1) $k$ PSS signals can be transmitted in the same angular direction to capture higher energy in a reduced time, and 2) to transmit $k$ PSS signal into different angular direction to reduce the search delay (in this work we only discuss the latter.).

According to the proposed scheme, the reduction in the directional search delay with higher $B_{SC,PSS}$ compared to $B_{SC}$ can be given as
\begin{equation}
   t_{Del,Red} = \frac{t_{Del}}{k}
    \label{eq:DelRed}
\end{equation}
where $t_{Del,Red}$ is a reduced delay. 
Note that this reduced delay is similar as we get with for $k$ times higher (fixed) sub-carrier bandwidth for all signaling. %, i.e., $t_{Del,Red} = t_{Del,kB_{SC}}$. 
However, with the proposed structure (assuming ABF), the BS can transmit k PSS signals in consecutive time instant in k different direction whereas with same $B_{SC}$ the BS can transmit PSS only in a single direction every $T_{PSS}$. We next discuss the EC vs $k$ relationship.

\begin{figure}
    \centering
    \psfrag{EC}{\back\back\back\back\back\back\back \ Energy Consumption (Joules)}
    \psfrag{BSc = 250 KHz}{$B_{SC} = 250$ KHz}
    \psfrag{k}{$k$}
    \psfrag{With Same Bsc}{\scriptsize $B_{SC} = kB_{SC}$}
    \psfrag{with Proposed Structure}{\scriptsize $B_{SC,PSS} = kB_{SC}$}
     \scalebox{0.9}[0.65]{\includegraphics[width=\columnwidth]{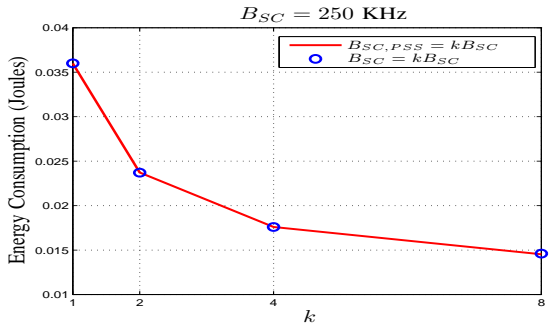}} \vspace{-4mm}
    \caption{Proposed slot structure for PSS.}
    \label{fig:ECvsk}
\end{figure}

Figure \ref{fig:ECvsk} shows the EC for different values of $k$.
The plot shows that the EC reduces with an increase in $k$, and therefore the propose signaling structure indeed reduces the EC. 
Moreover, the reduction in EC is similar to what can be achieved by using a fixed higher bandwidth sub-carrier for all signaling, as total time delay for both cases is similar. 
Therefore, the proposed structure with higher $B_{SC,PSS}$ can achieve a low EC (similar to what can be achieved with a higher $B_{SC}$), without increasing $B_{SC}$ for other signaling (which allows ADC to operate at lower sampling rate especially during data communication which reduces the power consumption). 
%However, in the proposed scheme $k$ PSS transmits in a burst, whereas in a fixed higher frequency sub-carrier scenario the PSS is transmitted in each direction every $T_{PSS}$.

%In summary, DBF may be a preferable choice for both the
%nCI and CID scenario as it has a minimum initial search delay
%and energy consumption. Moreover, ABF in CInD scenario has
%a minimum EC but its detection performance degradation in
%the presence of erroneous CI may not make it a preferable
%choice for ICD.

%%\enlargethispage{-56mm}
%
\section{Conclusions}
\label{sec:concl}
We studied the directional initial cell discovery problem for mmW 5G cellular networks in the context of an energy consumption comparison of ABF, DBF, HBF and PSN schemes.
We considered both context information (CInD, CID) and non context information (nCI) based scenario.
We showed that the perception regarding the higher energy consumption of DBF is not true during the directional initial cell discovery phase.
Rather, DBF has a lower energy consumption than ABF and HBF with much lower angular search delay for both CID (i.e., context information with delay) and nCI scenarios. 
Moreover, the recently proposed PSN architecture has a lower energy consumption than other beamforming schemes at higher sub-carrier bandwidth for nCI scenario, whereas ABF has minimum energy consumption in the CInD (i.e., CI with no delay) scenario.
We also showed that energy consumption results presented in this work (especially related to ABF, DBF and HBF) are valid irrespective to the number of ADC bits.
Finally, we proposed a new signaling structure which utilize a relatively higher bandwidth sub-carrier for PSS compared to other signaling which reduces the latency and energy consumption during initial cell discovery.

In the future, we will study the bandwidth vs noise power trade-off and identify the regimes where the energy consumption and the detection probability is maximized.

\bibliographystyle{IEEEtran}
\bibliography{IEEEabrv,biblio,refen}

\end{document}